\documentclass[12pt,letterpaper]{article}
\usepackage{setspace}
\usepackage{amsmath}
\usepackage{graphicx}

\begin{document}

\begin{center}
\textbf{Single-image diffusion coefficient measurements of proteins  in free solution}\vspace{3mm}\\
Shannon Kian Zareh$^{1}$, Michael C. DeSantis$^1$,  Jonathan Kessler$^1$, Je-Luen Li$^2$, \& Y. M. Wang$^1$\\
$^1$Department of Physics, Washington University in St. Louis, Saint Louis, MO 63130, USA\\$^2$D. E. Shaw Research, New York, NY 10036, USA\end{center}

\textbf{Abstract:} Diffusion coefficient measurements are important for many biological and material investigations, such as particle dynamics, kinetics, and size determinations.  Amongst current measurement methods, single particle tracking (SPT) offers the unique capability of  providing location and diffusion information of a molecule simultaneously while using only femptomoles of sample.  However, the temporal resolution of SPT is limited to seconds for single-color labeled samples.  By directly imaging three dimensional (3D) diffusing fluorescent proteins and studying the widths of their intensity profiles, we determine the proteins' diffusion coefficients using single protein images of sub-millisecond exposure times.  This simple method improves the temporal resolution of diffusion coefficient measurements to sub-millisecond, and can be readily applied to a range of particle sizes in SPT investigations and applications where diffusion coefficient measurements are needed, such as reaction kinetics and particle size determinations.\vspace{2mm}

Determination of particles' diffusion coefficients is important for many biological and material applications, such as single-molecule dynamics studies \cite{Wang2006,Greene2008,Moerner2010}, biolochemical and pharmaceutical reaction kinetics studies \cite{Stoeckli1998,Brock2009}, and particle size and shape determinations \cite{Niemeyer2003}.  Amongst current methods for measuring diffusion coefficients, which include nuclear magnetic resonance imaging (NMR) \cite{NMR2005}, dynamic light scattering (DLS) \cite{DLS2000}, fluorescence correlation spectroscopy (FCS) \cite{Webb1972,Nienhaus2004,Schwille2008}, and fluorescence recovery after photobleaching (FRAP) \cite{Wiersma2001}, single particle tracking (SPT) offers the unique capability of simultaneous location and diffusion coefficient determination.  This is essential for molecular mechanism investigations in heterogeneous environments, such as inside a cell's cytoplasm \cite{Xie2006}, flagella \cite{Nachury2010_2}, and membrane \cite{Spector2007} \textit{in vivo}, and on DNA molecules \cite{Wang2006} \textit{in vitro}.  Due to this capability, and the additional advantage that SPT experiments require less than femtomoles of sample, SPT can be a powerful tool in diffusion coefficient measurements applicable to a large range of biological investigations \textit{in vitro} and \textit{in vivo} where supplies are scarce.  

However, the drawback of using SPT for diffusion coefficient measurements is the low temporal resolution.  In single-molecule fluorescence imaging studies, stationary or slowly moving (relative to the data-acquisition timescales) single molecule intensity profiles are called point spread functions (PSF), and are fit to Gaussian functions to determine the molecules' localization information.  The centroid of the Gaussian function determines the lateral location of the molecule at the time of imaging, and the standard deviation (SD) determines the axial location.  In SPT diffusion coefficient measurements, consecutive locations of a single fluorophore are measured and from mean square displacement analysis of the particle's single trajectories, diffusion coefficients are obtained \cite{Wang2006,Xie2006,Qian1991}.  This method requires at least 20 consecutive location measurements for each single trajectory.  With the current single-photon camera imaging rate of approximately 100 frames/sec for a finite-sized imaging area, 0.2 sec is required and 3D diffusion coefficient $D_3$ measurements up to order $10^5$ nm$^2$/s have been reported \cite{Gratton2005}.  This requirement of 0.2 sec is, however, too long for diffusion coefficient measurements of fast-moving molecules, such as nanometer-sized proteins that diffuse beyond the typical imaging depth of $\sim$ 400 nm of single-molecule imaging microscope setups in less than 1 ms (a typical 5 nm protein has $D_3 \approx 10^8$ nm$^2$/s and diffuses $\sqrt{2D_3t} \approx 447$ nm in 1 ms).  A recent SPT method measures $D_3$ up to 1.7 $\times 10^7$ nm$^2$/s by labeling the particles with two colors \cite{Schmidt2011}; however, multi-color-labeling may not be feasible for many biological particles of interest and thus restricts the applicability of the method.

A SPT method that can determine 3D diffusion coefficients of single-colored nanometer-sized biological entities in their native environment is highly desirable for \textit{in vivo} and \textit{in vitro} studies.  In order to capture the molecule within the microscope's imaging depth, the imaging time needs to be less than 1 ms.   Here we report a novel method that determines the diffusion coefficient of nanometer-sized Brownian molecules from the SD values of the molecules' intensity profiles using sub-millisecond exposure times.  This is a single-image molecular analysis (SIMA) study of dynamic molecules and is an extension of our previous stationary molecule investigations \cite{Wang2010_2}.  In this study we used eGFP as the nanometer-sized fluorescent molecule for measurements and analyses.

Since the imaging times in our method are less than 1 ms, the temporal resolution of diffusion coefficient measurements is improved by at least 1000-fold over the minutes-long FCS method (multiple measurements each of order 20-second long),  200-fold over the 0.2-sec long centroid SPT method, 50-fold over the typically 50-ms long FRAP method, and 10-fold over the two-color SPT method.  Furthermore, the improvement in temporal resolution is achieved without compromising the precision of the $D_3$ measurements, and the single-image nature of the method avoids the photobleaching and limited lifetime photon problems associated with single-molecule fluorescence imaging studies.  Below we show our measurement method that relates the SD of a 3D freely diffusing protein's intensity profile to its diffusion coefficient $D_3$.  A prior study has used a similar concept to relate slow 2D diffusion coefficients (up to 1.1 $\times$ 10$^6$ nm$^2$/s) to a fluorophore's spot sizes \cite{Borczyskowski2002}; here we extend the study to fast 3D diffusion.
 
\begin{figure}
\includegraphics[width=5in]{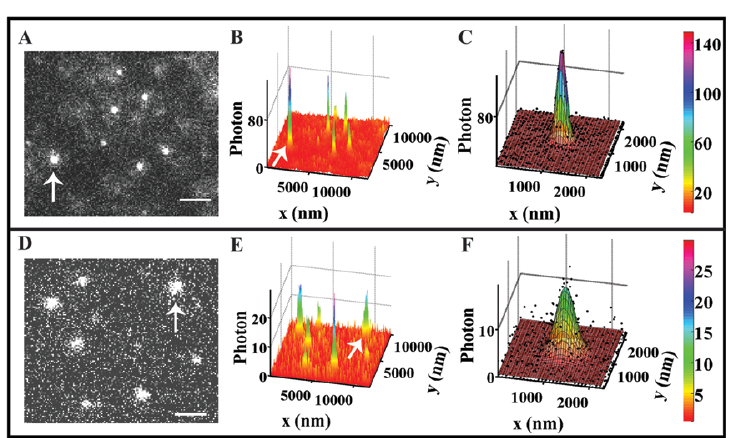}
\caption{Comparing stationary to diffusing eGFP molecules.  (A) An image of stationary eGFP molecules adsorbed on a fused-silica surface.  Five of the seven molecules have signal-to-noise ratio (SNR) $>$ 2.5.  (B) Intensity profiles of the stationary eGFP molecules in (A) in photon counts.  (C) Intensity profile (dots) and Gaussian fit (mesh) to the stationary eGFP molecule denoted by arrow in (A) and (B).  For this molecule, the SNR is 9.8, $s_x$ = 107.2 nm, and $s_y$ = 107.9 nm.  (D) Diffusing eGFP molecules near a reflective hydrophilic fused-silica surface at 1 ms exposure time.  Six of the eight molecules have a SNR $>$ 2.5.  The scale bars for (A) and (D) are 2 $\mu$m.  (E) Intensity profiles of the diffusing eGFP molecules in (D).  (F) Intensity profile (dots) and Gaussian fit (mesh) to the diffusing eGFP molecule denoted by arrow in (D) and (E).  For this molecule, the SNR is 3.5, $s_x$ = 202.2 nm, and $s_y$ = 192.4 nm.  It is clear that the intensity profiles of diffusing molecules are wider (or have larger SDs) than that of stationary molecules.}
\label{Fig1}
\end{figure}

\begin{figure}
\includegraphics[width=5in]{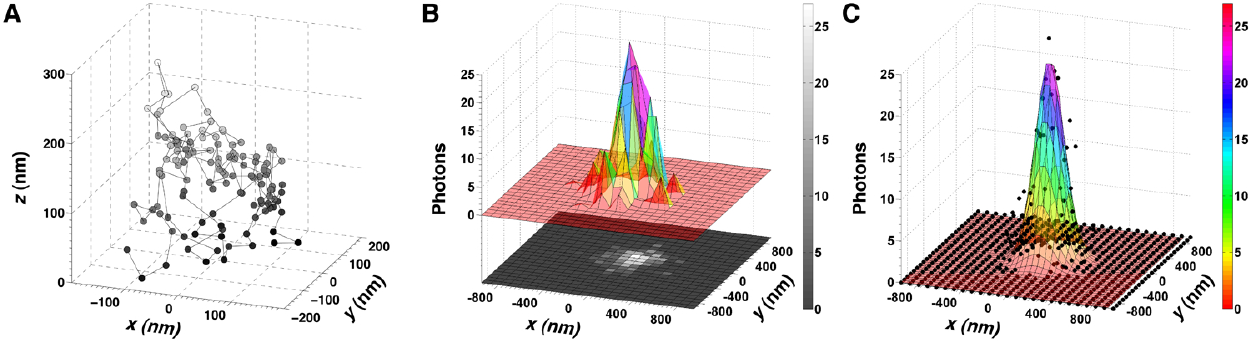}
\caption{Simulated image formation and analysis process of a diffusing eGFP molecule.  (A) Trajectory of a diffusing eGFP molecule in free solution under TIRF (total internal reflection fluorescence) evanescent excitation at the exposure time of 0.6 ms.  The data is gray-scaled to correspond to the particle's axial locations (SOM Text).  (B) The emitted photons from the trajectory form an intensity profile (colored plot), which is then projected onto a 2D camera screen (black and white image).  (C) Gaussian fit (mesh) to the intensity profile of the diffusing eGFP (dots), where $s_x$ = 119.4 nm, and $s_y$ = 142.2 nm.}
\label{Fig2}
\end{figure}

Figures \ref{Fig1} and \ref{Fig2} illustrate the principle of this method.  In a finite exposure time, the intensity profile of a moving molecule is wider (or more blurry) compared to that of an immobile molecule.  Figure \ref{Fig1}A shows a 30 ms frame image of stationary eGFP molecules adsorbed on a fused-silica surface, and Fig. \ref{Fig1}D shows a 1 ms frame image of diffusing eGFP molecules near a hydrophilic fused-silica surface \cite{Wang2011_2}.  These figures clearly show that diffusing molecules images are blurry compared to that of immobile molecules.  In Figs. \ref{Fig1}B and \ref{Fig1}E the intensity profiles of the stationary and diffusing eGFP molecules are plotted, and in Figs. \ref{Fig1}C and \ref{Fig1}F, the respectively selected intensity profiles are fitted to Gaussian functions.  While both intensity profiles fit well to a Gaussian function, the width (or SD) of the diffusing protein's intensity profile is larger than that of the stationary protein.  

In general, the final image of a diffusing molecule, like those in Fig. \ref{Fig1}, is the sum of the emitted photons along its diffusion trajectory projected onto a 2D imaging screen during the exposure time.   Figure \ref{Fig2}A shows a simulated eGFP diffusion trajectory at 0.6 ms exposure time using 0.005 ms steps for clarity.  The data are gray-scaled to correspond to the particle's axial locations (SOM text).  The emitted photons, after photon-to-camera count conversion, were projected onto a 2D imaging screen and binned into our camera pixels each of 79 $\times$ 79 nm$^2$ in size (Fig. \ref{Fig2}B; gray image) and the corresponding diffusing eGFP PSF intensity profile was formed in the colored image above.  The total photon count of this image was 414.  The 2D Gaussian fit to the diffusing eGFP intensity profile is shown in Fig. \ref{Fig2}C, yielding SD values in the $x$- and $y$-directions, $s_x$ and $s_y$, respectively.   $s_{x,y}$ values presented in this article are results from fitting to these experimental and simulated PSF data, and they were used to quantify the blur of diffusing eGFP molecules, and consequently, the diffusion coefficient $D_3$.   

In order to determine $D_3$ from diffusing fluorophore images, we performed (i) experimental measurements, (ii) analytical calculations, and (iii) simulations and obtained SDs of diffusing eGFP intensity profiles at 0.3 to 1 ms exposure times.  Below we show that when the experimental results were checked against theoretical calculation and numerical simulation results in Fig. \ref{Fig5}A, the good agreement validates our method of measuring nanometer-sized fluorophore diffusion coefficients.

\begin{figure}
\includegraphics[width=5in]{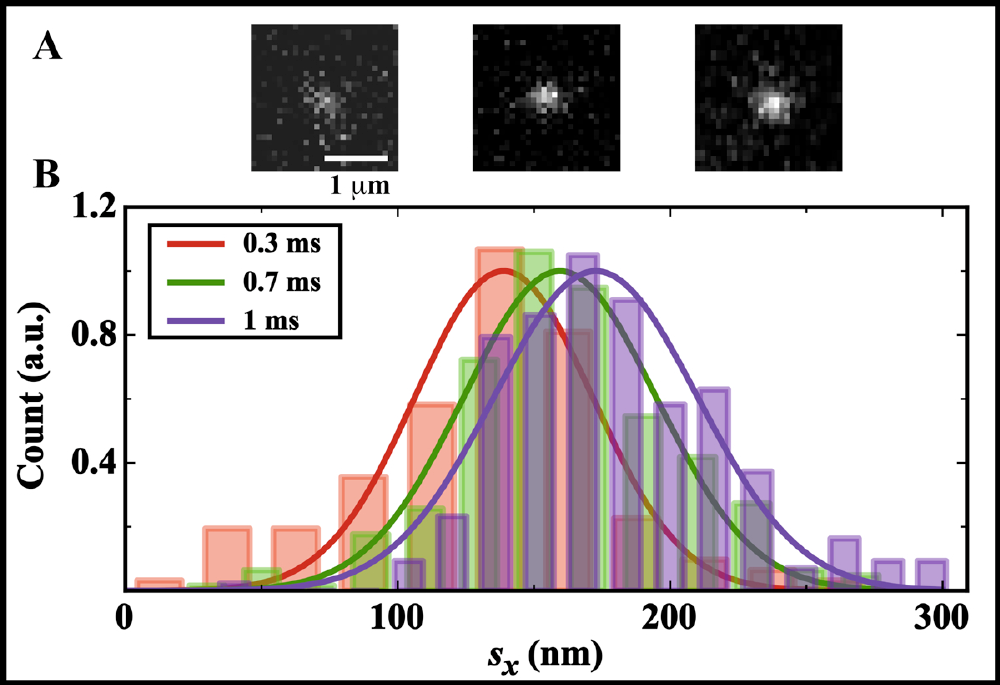}
\caption{Diffusing eGFP images and intensity profile SD distributions at different exposure times.  (A) Three representative images showing diffusing eGFP molecules at exposure times of 0.3, 0.7, and 1 ms.  The intensity profile SD values increase with the exposure time.  The scale bar is 1 $\mu$m.  (B) EGFP intensity profile SD distributions (normalized by counts for comparison) at the three aforementioned exposure times, showing increasing values of 136.8 $\pm$ 27.7 (mean $\pm$ SD), 159.0 $\pm$ 32.24, and 172.1 $\pm$ 34.8 nm, respectively.}
\label{Fig3}
\end{figure}
\begin{figure}
\includegraphics[width=4in]{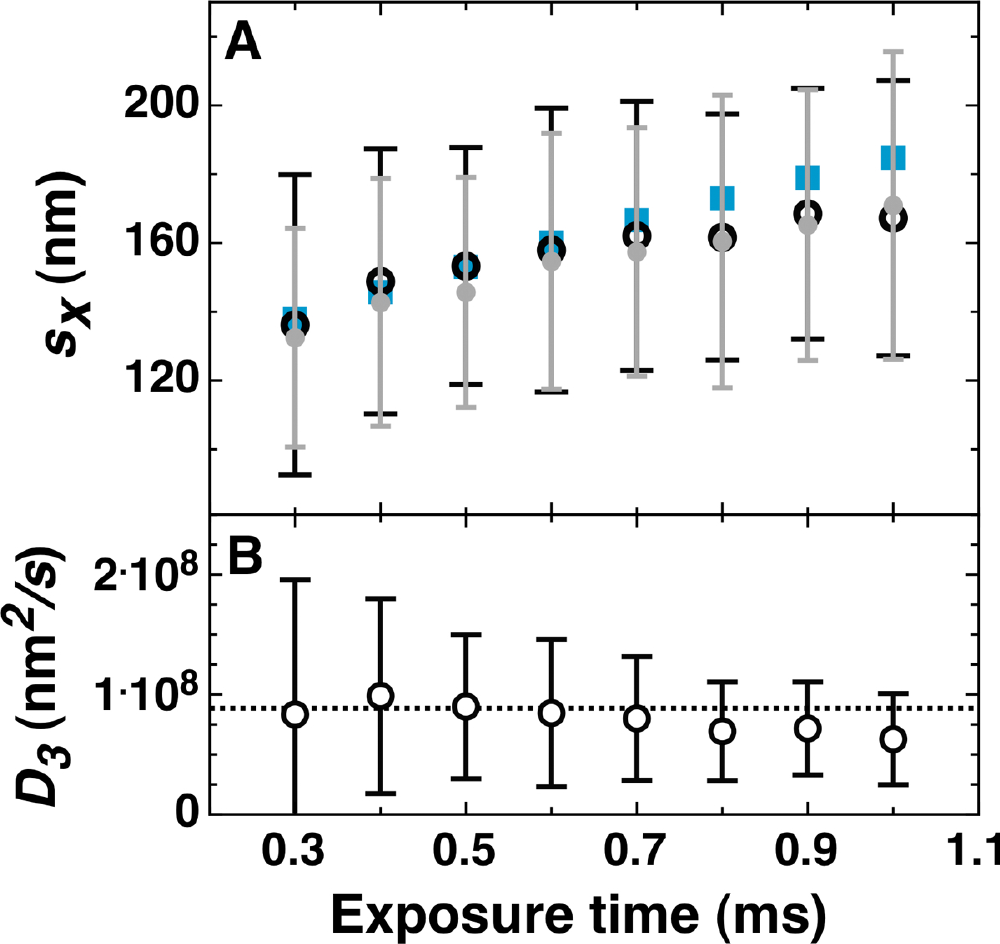}
\caption{Comparing $s_x$ and $D_3$ results.  (A) Experimental (circles), simulation (disks), and theoretical calculation (squares) measurements of diffusing eGFP intensity profiles' mean $s_x$ vs $t$.  In the experimental and simulation results, the error bars are the SDs of the $s_x$ distributions.  (B) Experimental $D_3$ calculated from Eq. \ref{D3}.  The error bars are $\Delta{D_3}$ calculated using Eq. \ref{error}; the dashed line is the FCS-determined eGFP $D_3$ of 8.86 $\times 10^7$ nm$^2$/s for comparison.}
\label{Fig5}
\end{figure}

In experimental measurements, Fig. \ref{Fig3}A shows representative eGFP images (chosen such that the molecule's respective $s_x$ values were within $\pm$ 5 nm of the means to the respective diffusing eGFP intensity profile SD distributions in Fig. \ref{Fig3}B) at 0.3, 0.7, and 1 ms exposure times.  As expected, SD values of these respective single diffusing eGFP molecules increase from 136.4 to 160.9 and 175.5 nm, validating that SD provides a quantitive measure for the motion-induced blurriness of single fluorophore images.  The increasing mean eGFP SD values with exposure time are plotted in Fig. \ref{Fig5}A, where the error bars are the SDs of the eGFP intensity profile SD distributions in Fig. \ref{Fig3}B.


In analytical calculations, we deduce an expression relating a diffusing eGFP's SD to $D_3$.  The study involves first projecting the eGFP PSFs at all focal depth onto a 2D imaging screen, forming an axial-direction-projected PSF $f(x,y)$, and then convolving this projected PSF with the lateral location distribution of the molecule in a trajectory, which we define as a pathway distribution function (PWDF$_{x,y}$ for the lateral directions) $g(x,y)$: 
\begin{equation}
I(x,y) {\propto} f(x,y){\ast}g(x,y).
\label{convolution1}
\end{equation}

In SOM we show that both $f(x,y)$ and $g(x,y)$ can be approximated well by Gaussian function for sufficiently low exposure times,  and their convolution is another Gaussian function with a variance equal to the sum of the two variances.  Therefore, the final projected intensity profile of a 3D diffusing molecule is a Gaussian function with SD being
\begin{equation}
s_{x,y} = \sqrt {s_0^{\prime{2}}+A_{x,y}\cdot2D_3t},
\label{convolution3}
\end{equation}
where $s_0^{\prime}(t) = \sqrt{111^2 + 0.0634D_3t} \approx \sqrt{s_0^2 + 0.0634D_3t}$ nm is the SD of the axial-direction-projected PSFs for our experimental parameters, and $A_{x,y}\cdot2D_3t$ is the variance of PWDF$_{x,y}$s with $A_{x,y} = 0.0926$. This relation enables the determination of $D_3$ from the SD of a single-molecule's intensity profile and the exposure time as
\begin{equation}
D_3 = \frac{{s_{x,y}}^2-s_0^2}{(2A_{x,y}+0.0634)t}.
\label{D3}
\end{equation}  

Using FCS-determined eGFP $D_3 = 8.86 \times 10^7$ nm$^2$/s in Eq. \ref{convolution3}, the analytical eGFP $s_{x}$ results are plotted in Fig. \ref{Fig5}A, showing excellent agreement with the experimental $s_{x}$ results within 0.7 ms.  Note that $s_{x}$ starts to deviate from the experimental  results at $t >$ 0.8 ms; this is because the exposure time begins to approach the diffraction-limit-determined value for eGFP for this study (SOM text).
 
We have also performed simulations of diffusing eGFP intensity profiles (as shown in Fig. \ref{Fig2}) using the FCS-determined $D_3$.  Figure \ref{Fig5}A juxtaposes the simulated diffusing eGFP SD results with the experimental results; the two mean values and error bars agree at all exposure times (Fig. S7 
compares the results at $t$ = 0.6 ms).  
 
To determine the precision of the measured $D_3$ from single eGFP images, we performed error propagation analysis of $D_{3}(s_{x,y})$ using Eq. \ref{D3}:  
\begin{equation}
\Delta{D_3} = \frac{s_{x,y}}{(A_{x,y}+0.032)t}\Delta{s_{x,y}},
\label{error}
\end{equation}
where $\Delta{s_{x,y}}$ is the SD measurement precision of the single fluorophore's intensity profile (the experimental error bars in Fig. \ref{Fig5}A, also Ref. \cite{Wang2010}).  Figure \ref{Fig5}B compares the experimentally determined $D_3$ and $\Delta{D_3}$ from single diffusing eGFP image SD measurements to the FCS-determined eGFP $D_3$ = 8.86 nm$^2$/s, showing agreement. 

At 0.7 ms,  $\Delta{D_3} = $ 5.2 $\times 10^7$ nm$^2$/s for a single eGFP image using both the statistically independent mean $s_x$ and $s_y$ values of $\langle{s_{x,y}}\rangle$ = 162.1 nm and $\Delta{s_{x,y}}$ = 39.2 nm.  It is 57$\%$ of the eGFP $D_3$ value of 8.86 $\times 10^7$ nm$^2$/s.  Since there are order 30 molecules in a typical frame image of less than 1 ms exposure time, the precision to $D_3$ measurement further improves by $\sqrt{30}$ times to 10$\%$, comparable to the precision of FCS $D_3$ measurements \cite{Schwille2008}.  In spatially restrictive situations, such as \textit{in vivo} imaging in typically micron-sized cells, where only one image can be obtained at a time, repeated single-image measurements will enable precise $D_3$ determination.

Although this study focuses on fast diffusion of nanometer-sized proteins in free solution with $D_3 > 5 \times 10^7$ nm$^2$/s, the methodology applies to 3D diffusion of all rates.  When diffusion coefficients are low for large particles, in a crowded environment, or in viscous solvents (such as in cells \cite{Verkman1997} or glycerol), the molecule's intensity profile will be more localized.  Consequently, longer exposure times should be used to observe noticeable changes to the  SD from the stationary values.  SOM text explains the procedure in determining the appropriate exposure times for a particle of unknown $D_3$.  

Molecules' movement can deviate from a 3D unbiased Brownian motion, examples include directional motion and diffusion with drift.  For these alternative motions, $s_0^{\prime}$ and PWDF$_{x,y,z}$ should be determined before convolving the axial-direction-projected PSF with PWDF$_{x,y}$ for the final intensity profile.  As long as the mean numerical SD of locations in the molecule's trajectory is less than half of the diffraction limit at the exposure time, the projected convolved image of the molecule will be a unimodal intensity profile that can be fitted to a Gaussian function, and the resulting $s_{x,y}$ will provide information on the molecule's dynamics.  

In summary, we present a new single-molecule fluorescence image analysis method that measures fast diffusion coefficients with high precision.  The experimental setup and data analysis are simple while using standard microscopy imaging systems, and the method is applicable to a wide range of diffusion coefficient measurements with greatly improved temporal resolution.  Applications in basic research and pharmaceutical investigations such as fast drug screening can be envisioned.


\newpage
\bibliographystyle{Science}

\pagebreak
\clearpage
\newpage
\newpage
\newpage


\hspace{-6mm}\textbf{Supporting Online Material}\\
\textbf{Materials and methods}\\
\textit{Sample preparation and imaging}\\
The eGFP molecules (4999-100, BioVision, Mountain View, CA) were diluted in 0.5X TBE buffer (45 mM Tris, 45 mM Boric Acid, 1 mM EDTA, pH 8.0) to 0.03 nM.  For the stationary eGFP studies, manufacturer pre-cleaned fused-silica chips (6W675-575 20C, Hoya Corporation USA, San Jose, CA) were used, where isolated eGFP molecules were adsorbed to surfaces at low concentration.  For the diffusing eGFP studies, the manufacturer pre-cleaned fused-silica chips were treated using oxygen plasma for three minutes, rendering it hydrophilic to prevent eGFP adsorption \cite{Wang2011_2}.  The hydrophilic fused-silica surface can be considered ballistic for the diffusing eGFP molecules in our experiments and simulations.  For both studies, a protein solution of 5 $\mu$L was sandwiched between the fused-silica surface and an oxygen-plasma-cleaned coverslip (2.2 $\times$ 2.2 cm$^2$), resulting in a 10.5 $\mu$m thick water layer.  The coverslip edges were then sealed with nail polish.  

Single-molecule imaging was performed using a Nikon Eclipse TE2000-S inverted microscope (Nikon, Melville, NY) in combination with a Nikon 100X objective (Nikon, 1.49 $N.A.$, oil immersion).  Samples were excited by prism-type total internal reflection fluorescence (TIRF) microscopy with a linearly polarized 488 nm laser line (I70C-SPECTRUM Argon/Krypton laser, Coherent Inc., Santa Clara, CA) focused on a 40 $\times$ 20 $\mu$m$^2$ region.  The 488 nm line was filtered from the multiline laser emission using a polychromatic acousto-optic filters (48062 PCAOM model, NEOS Technologies, Melbourne, FL).  The laser excitation was pulsed with illumination interval of 30 ms for the stationary eGFP molecules in Fig. 1 
and Fig. S\ref{FigS6}, and between 0.3 ms and 1 ms for the diffusing eGFP molecules. The excitation intensities were 2.7 and 3.2 kW/cm$^2$ for the respective stationary eGFP molecules, and 37.5 kW/cm$^2$ for the diffusing molecules.  Images were captured by an iXon back-illuminated electron multiplying charge coupled device (EMCCD) camera (DV897ECS-BV, Andor Technology, Belfast, Northern Ireland).  An additional 2X expansion lens was placed before the EMCCD, producing a pixel size of 79 nm.  The excitation filter was 488 nm/10 nm, and the emission filter was 525 nm/50 nm. \vspace{2mm}\\
\textit{Data acquisition and selection}\\
Movies were obtained by synchronizing the onset of camera exposure with laser illumination for different intervals.  The maximum gain level of the camera was used and the data acquisition rate was 1 MHz pixels/sec ($\approx$ 3.3 frames/sec).  Single-molecule images were checked such that there were no saturations in the intensity profiles.  For the defocusing analysis of stationary eGFP molecules, 21 $\times$ 21 pixel boxes centered at the molecule were selected by hand using \textsc{ImageJ} (NIH, Bethesda, MD), and the intensity values were used for the 2D Gaussian fitting.    For the diffusing eGFP molecule movies, all visible diffusing eGFP intensity profiles in the peak laser excitation region of 10 $\times$ 10 $\mu$m$^2$ were selected by hand using 39 $\times$ 39 pixel boxes centered at the molecule.  The center 25 $\times$ 25 pixels of the boxes  were used for 2D Gaussian fitting, and the peripheral pixels were used for experimental background analysis.  

Before analysis, the camera's intensity count at each pixel in an image was converted into photon count by using the camera-to-photon count conversion factor calibrated the same day of the measurement as described in our previous article \cite{Wang2010}.  The number of detected photons in an image was obtained by subtracting the total photon count of the BG from the total photon count of the image.  The eGFP intensity profiles were fit to a 2D Gaussian function in order to obtain the SD values of the molecule:
\begin{equation}
f(x,y)=f_0  \exp{\left[-\frac{(x-x_0)^2}{2 s^2_x} - \frac{(y-y_0)^2}{2 s^2_y}\right]}+ \langle b \rangle,
\label{Gaussian}
\end{equation}
where $f_0$ is the multiplication factor, $s_x$ and $s_y$ are SDs in the $x$- and $y$- directions, respectively, $x_0$ and $y_0$ are the centroid location of the molecule, and $\langle{b}\rangle$ is the mean background offset in photons.  

For the defocusing eGFP analysis, we selected 17 adsorbed eGFP molecules with a minimum photon count of 229 and signal-to-noise ratios (SNR, $I_0/\sqrt{I_0+\sigma_b^2}$) of higher than 3.75, where $I_0$ is the peak PSF photon count (after subtracting the mean background offset $\langle{b}\rangle$) and $\sigma_b^2$ is the background variance in photons.  For the diffusing eGFP molecules, we used a SNR of 2.5 as a selection criterion, and PSFs with photon counts less than 50 were not used in the analysis because they would be invisible in experimental data.  At each exposure time, we acquired 1600 data points from 4 movies recorded at different regions of the imaging chip; the number of diffusing eGFP data used for the experimental analysis that satisfied the SNR criteria are 419 to 1066 for the 0.3 ms to 1 ms exposure times, respectively.  \vspace{2mm}\\
\textit{Analytical expression of diffusing eGFP $D_3(s_x)$.}\\
In this section we decompose an eGFP's 3D diffusion process into two components for $s_x$ and $D_3$ calculation:  a 1D diffusion along the axial direction and a 2D diffusion in the lateral directions.  

It is known that as the defocusing distance between the fluorophore and the focal plane increases the PSF's SD increases as well. Consequently, calculation of the intensity profile necessitates integrating over all axial locations the molecule may have traveled during the exposure time to obtain an axial-direction-projected PSF, $f(x,y)$.  As diffusion in the lateral and axial dimensions are statistically independent of each other, we choose to perform this integration prior to convolving the resulting PSF with PWDF$_{x,y}$ in the lateral dimensions to obtain the final projected 2D intensity profile of the 3D diffusing molecule on an imaging screen.   

In the axial direction, the axial-direction-projected PSF is computed by numerically integrating defocused PSFs through $z$ for all pixelated $x,y$ values (sufficient to contain all defocused PSFs)
\begin{equation}
\int_0^{400}C(z)\exp\left(-\left[\frac{x^2}{2s_x(z)^2}+\frac{y^2}{2s_y(z)^2}\right]\right){\exp}\left(-\left[\frac{(z-\langle{z_0}\rangle)^2}{2A_z\cdot2D_3t}+\frac{z}{d}\right]\right)dz,
\label{Defocus}
\vspace{-0mm}
\end{equation}
where $C(z)$ and $s_{x,y}(z)$ are the amplitude and SDs of our imaged, defocused eGFP Gaussian PSFs (SOM text), respectively, $\langle{z_0}\rangle$ and $A_z\cdot2D_3t$ are the mean and variance of diffusing eGFPs' Gaussian PWDF$_z$s (SOM text), $\exp(-z/d)$ describes the decaying TIRF evanescent excitation intensity, and the range for the $z$-integration is the imaging depth of 0 nm to 400 nm measured from the focal point at the fused-silica surface.  The resulting axial-direction-projected PSF $f(x,y)$ remains Gaussian, and the SD is a function of the exposure time $t$ as $s_0^{\prime}(t) = \sqrt{111^2 + 0.0634D_3t}$ nm.  

In the lateral directions, the intensity profile $I(x,y)$ of a diffusing eGFP's image is the convolution of the axial-direction-projected eGFP PSF $f(x,y)$ with the PWDF$_{x,y}$, $g(x,y)$ as 
\begin{equation}
I(x,y) {\propto} f(x,y){\ast}g(x,y).
\label{convolution1}
\end{equation}
 
\begin{figure}
\includegraphics[width=5in]{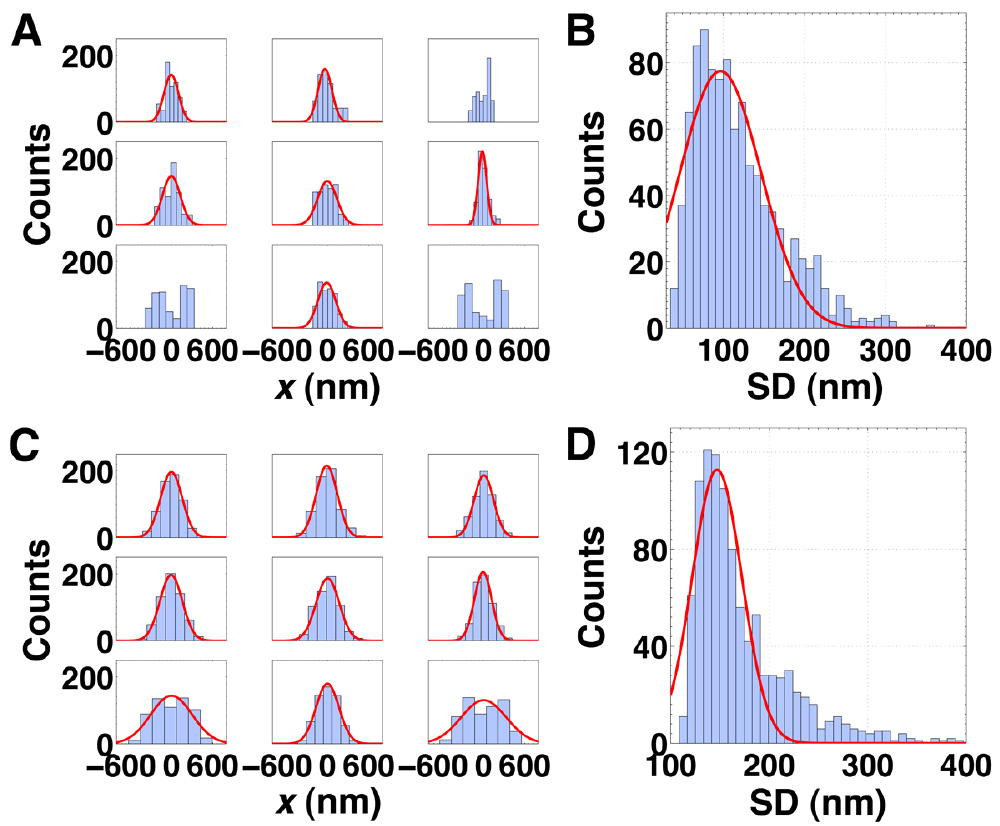}
\caption{Study of the eGFP lateral PWDF$_{x}$s and their convolution with PSFs.  (A) Nine random eGFP PWDF$_{x}$s at 0.6 ms exposure time and Gaussian fits to the unimodal distributions with $R^2 >$ 0.8.  (B) The distribution of 1000 PWDF$_x$ SDs, fitted with a Gaussian.  (C) The 9 PWDF$_x$s in (A)  convolved with eGFP PSFs at focus with $s_0$ = 108.2 nm.  (D) The SD distribution of 1000 PWDF$_x$ convolved eGFP PSFs at focus and its Gaussian fit.}
\label{Fig4}
\end{figure}

We numerically calculate $g(x,y)$ of a freely diffusing eGFP particle by simulations.  Figure S\ref{Fig4}A shows 9 random PWDF$_x$s at exposure time $t = 0.6$ ms.  Six of the nine PWDF$_x$s have one peak (uni-peaked or unimodal) and can be fitted to a Gaussian function with $R^2 >$ 0.8.  Figure S\ref{Fig4}B shows the SD distribution of PWDF$_x$s, combining the Gaussian fitted SD values for the uni-peaked PWDF$_x$s and the numerical particle location distribution SD values for the double-peaked PWDF$_x$s; the mean is 96.8 nm.  Figure S\ref{Fig4}C shows that when the 9 PWDF$_x$s in Fig. S\ref{Fig4}A  are convolved with single-eGFP PSFs at focus with $s_0$ = 108.2 nm, all convolved PWDF$_x$s fit well to a Gaussian function, and the mean of the SD distribution is 147.1 nm.  Therefore, although not all PWDF$_x$s are uni-peaked, taken over all, we can view PWDF$_x$s as Gaussian functions with an average $t$-dependent SD value of $\sqrt{A_{x}{\cdot}2D_3t}$.  For the 0.6 ms exposure time data, $A_{x}$ = 0.0882.  We found $A_{x}$ to be insensitive to exposure times below 1 ms (mean $A_{x}$ is 0.0926; data not shown).
 
Given that $f(x,y)$ (at focus and the axial-direction-projected) and $g(x,y)$s are both Gaussian functions,  in the lateral directions, their convolution can be described by another Gaussian function with a variance equal to the sum of the two variances.  Using the focused eGFP PSFs with $s_0$ = 108.2 nm and PWDF$_x$ at 0.6 ms, $s_{x,2D} = \sqrt {s_0^2+A_{x}\cdot2D_3t} = \sqrt{108.2^2+96.8^2}$ nm = 145.2 nm, very close to the mean SD value of the above PSF-convolved-PWDF$_x$s of 147.1 nm.

Finally, we can calculate the intensity profiles' SDs of freely 3D diffusing molecules.  Since we have observed that both the axial-direction-projected PSFs (data not shown) and the lateral PWDF$_{x,y}$s are Gaussian,  the final projected intensity profiles' SD of diffusing molecules is
\begin{equation}
s_{x,y} = \sqrt {s_0^{\prime{2}}+A_{x,y}\cdot2D_3t}.
\label{convolution3}
\end{equation}
This relation enables the determination of $D_3$ from the SD of a single-molecule's intensity profile and the exposure time.\vspace{2mm}\\  
\textit{Diffusing eGFP Simulations}\\
We simulated 3D Brownian diffusion eGFP trajectories at a range of exposure times using FCS-determined eGFP $D_3$ = 8.86 $\times$ 10$^7$ nm$^2$/s  and triplet state statistics.  The starting locations of the trajectories followed the distribution function described in the SOM Text.  The step sizes in the $x, y$, and $z$ directions were randomly selected from a Gaussian distribution with a mean of zero and SD of $\sqrt{2D_3t_0}$ with a step time $t_0 = 1$  $\mu$s.  Because of the reflective fused-silica-water interface, the simulated $z$ values were maintained above zero.  The number of steps in a simulation was $t/t_0$.  At each $x,y$ location in a trajectory, when the molecule was not in a triplet dark state, a Poisson distributed number of photons (described in SOM text) were drawn from a Gaussian PSF spatial distribution with a mean of zero and the corresponding SD value for the axial-location (SOM text). This relative displacement of the photons is added to the simulated $x,y$ location of the molecule, generating the actual $x,y$ location of the emitted photons at the simulation step.  


The simulated photons of each trajectory were binned into $50 \times 50$ pixels with a pixel size of 79 nm.  Then the photon count of each pixel was converted into the modified camera count using Eq. 4 in Ref. \cite{Wang2010} with the photon multiplication factor of the camera $M$ = 1 in order to include the camera count variance effect.  Random background photons at each pixel were generated using the corresponding experimental background distribution functions for the exposure time (described in Ref. \cite{Wang2010}).     The final intensity profiles were fit to a 2D Gaussian function to obtain the two SD values for the image.  For each SD datum of diffusing eGFP molecules in Fig. 4, 
1000 independent trajectories were simulated.  

\newpage
\hspace{-6mm}\textbf{SUPPORTING TEXT}\\
%
\textbf{Exposure time limit determination for $D_3$ measurements using Eq. 3
.}\vspace{0mm}\\
In Fig. 4
, the calculated SD starts to deviate from the experimental and simulation results at 0.8 ms.  This suggests the existence of an upper bound exposure time for our eGFP studies.  In order to determine the appropriate exposure time, we explain the origin of this deviation at long exposure times.  

The PWDF$_x$s in Fig. S\ref{Fig4}A show both unimodal and double-peaked patterns.  When convolved with PSFs, at short exposure times, both the unimodal and double-peaked PWDF$_x$s will produce unimodal intensity profiles suitable for Gaussian fitting; however, at long exposure times, the two-peaked PWDF$_x$s will yield a two-peaked intensity profile.  As the exposure time increases, both the fraction of two-peaked PWDF$_x$s and the peak separations increase, while Eq. \ref{Defocus} still assumes Gaussian fits for all PSF-convolved-PWDF$_x$s.  As a consequence, a deviation between the analytical and the experimental $s_x$ appears and increases with exposure time.   

The threshold exposure time for the onset of the deviation is determined as follows.  When two identical fluorophores are separated by more than the diffraction limit, the combined intensity profile is double-peaked \cite{Wang2010_2}.  However, when the concentration of fluorophores peaks in between the diffraction limit separation of the instrument, as for unimodal PWDF$_x$s, the combined intensity profile appear to be unimodal,  and can be fitted to a Gaussian function.  For the double-peaked PWDF$_x$s, which creates two clusters of fluorophores separated by the distance between the two peaks, this distance, which is approximately the numerical 2$\times$SD value of the molecule's location distribution, or $2\cdot \sqrt{A_{x}{\cdot}2D_3t}$ of the PWDF$_x$s, determines the exposure time limit for when the convolved intensity profiles become two-peaked.  When $2\cdot \sqrt{A_{x}{\cdot}2D_3t}$ is less the the diffraction limit of 217 nm, the convolved intensity profile is unimodal and can be approximated by a Gaussian function.  The threshold is crossed at the exposure time $t\approx 0.8$ ms, where the PWDF$_x$'s mean SD value is 113.6 nm.  


From the above analysis, for $D_3$ determination using Eq. 3 in the main text
, the upper bound exposure time can be determined by requiring $2\cdot\sqrt{A_{x}{\cdot}2D_3t}$ to be less than the diffraction limit separation of the imaging setup's emission wavelength and \textit{N.A}.\vspace{2mm}\\
\textbf{Choosing exposure times for a particle of unknown $D_3$.}\\
For particles of unknown $D_3$, exposure time can be scanned until the diffusing particle images are noticeabily larger than that of stationary particles, while remaining unimodal.  In this range of exposure times, we can use Eq. 3 
 for eGFP to calculate $D_3$ of this particle for the following reason: in Eq. 3, 
  since $s_{0}^{\prime}$ is calculated from integration that depends on $D_3t$ (SOM), and $A_{x,y}\cdot2D_3t$ varies with $D_3t$ only, at the appropriate exposure time, although the $D_3$ values of the particle and eGFP are different, the $D_3t$ values can be equivalent.  At these exposure times, Eq. 3 
 for eGFP is restored valid for the particle and can be used to determine the unknown $D_3$.\vspace{2mm}\\
\textbf{FCS  determination of the eGFP diffusion coefficient.}\\
In order to independently verify our experimentally determined eGFP PSF mean SD results (Eq. 3) 
; therefore also the $D_3$ result) by using theoretical calculations and simulations, we performed FCS $D_3$ measurements of eGFP (at the Washington University Fluorescence Correlation Spectroscopy and Confocal Imaging Facility in the Department of  Biochemistry and Molecular Biophysics).  

In FCS measurements, fluorescence from freely diffusing eGFP molecules at 3 nM concentration in 0.5X TBE buffer (pH 8) was measured.   An autocorrelation function was used to obtain the eGFP diffusion parameters \cite{Briddon2004},
\begin{equation}
G(t) = \frac{1}{N(1+\frac{\tau}{\tau_d})\sqrt{(1+\frac{\tau}{s^2\tau_d})}}{\left(\frac{1-F+Fexp(-\frac{\tau}{\tau_k})}{1-F}\right)} +1,
\label{FCS}
\end{equation}
where $\tau$ is the detection time, $N$ is the number of molecules in the detection radius $w$, $s$ is the structure parameter of the excitation beam focal region (the ratio of the beam radius in $z$ to the beam radius in $x$ and $y$), $\tau_d = \frac{w^2}{4D_3}$ is the molecule's diffusion time in the imaging area, $F $ is the fraction of molecules in triplet state, and $\tau_k$ is the triplet state lifetime.  

The excitation wavelength for the FCS measurement was 488 nm, and the emission photons went through a 505 -- 550 nm filter.  The excitation power was 76.4 kW/cm$^2$, which was comparable to our excitation power of 37.5 kW/cm$^2$ in the diffusing eGFP studies.  We used Alexa 488 with a known diffusion coefficient $D_3 = 4.35 \times 10^8$ nm$^2$/s \cite{Schwille2008} for calibration and obtained $w$ $\approx$ 250 nm.  For Alexa 488, $\tau_d = 35.6$ $\mu$s.  Figure S2 
shows the $G(t)$ vs. $t$ plot of the eGFP system, where $\tau_d$ was 174.8 $\mu$s.  Assuming a Gaussian detection volume and using a one-component fit, the best fit to Fig. S \ref{FCS} yields $F = 12.7$, $\tau_k = 3$ $\mu$s, and $s = 10$.  Using $\tau_d = \frac{w^2}{4D_3}$, we obtained eGFP $D_3 = 8.86 \times 10^7$ nm$^2$/s.   This value of eGFP $D_3$ is consistent with reported values \cite{Schwille2008}.\vspace{2mm}\\
\textbf{Refractive index mismatch corrections to eGFP intensity profiles.}\\
When a fluorescent molecule in water is imaged through a glass coverslip using a high \textit{N.A.} oil immersion objective, the refractive index mismatch between the water-based solvent and the glass coverslip changes the fluorophore's intensity profiles in two major ways \cite{Hell2006}: (i) Due to Snell's law of refraction, the actual axial location of the molecule (measured from the glass coverslip-water interface) is deeper than the apparent axial position of the molecule (defined by the depth in water where the imaged fluorophore's PSF amplitude is maximal). (ii) Due to spherical aberrations, if the focus is at the apparent position of the molecule (we define as $z$ = 0), the fluorophore's defocused intensity profile's SD vs. $z$ relation is asymmetric with respect to $z$ = 0 \cite{Hell2006}.  Figure S3A 
 shows the geometry of our setup,  where the direction of $z$ is positive towards the glass coverslip, opposite to that of Ref. \cite{Hell1999}.  In our prism-type-TIRF imaging setup, the eGFP molecules were adsorbed on or diffusing near the fused-silica surface in TBE buffer 10.5 $\mu$m from the coverslip-water interface.  We used fused-silica as the TIRF interface because of its low background noise and thus high SNR for the study when comparing to objective-TIRF imaging that uses glass coverslips as the TIRF interface, where the mean background noise level is at least 6 times higher than ours at comparable laser intensity and exposure times (data not shown).  

In order to obtain accurate defocusing eGFP intensity profile parameters for SD and $D_3$ calculations, we performed calculations and measurements of eGFP adsorbed on fused-silica surfaces.  In calculations, we obtained the defocused fluorophores' PSF using the diffraction integral analysis in Ref. \cite{Hell1999}, which has been used by other groups for mainly calculating the actual axial location of the imaged fluorophores \cite{ZhuangCylinder2008,Moerner2010}).  We assume the final PSF of a defocused fluorophore to be the average of 4 emission polarizations at 0, $\pi/4$, $\pi/2$, and $3\pi/4$, and the light intensity at the spherical wavefront of the point light emitter before reaching the objective was homogeneous.  Figures S3B and C 
 show the SD and amplitude of the calculated fluorophore PSFs using our imaging system's parameter of $N.A.$ = 1.49, water's refractive index of $n_1 = 1.34$, glass' refractive index of $n_2 = 1.515$, and an emission wavelength of 525 nm (plotted in blue).


The experimental measurements of defocused eGFP molecules were performed with the molecules adsorbed on fused silica surfaces and a focus-drive (H122, Prior Scientific Inc., Rockland, MA) moving one-way in 100 nm increments through the focal point.   The average SD and normalized amplitude of eGFP intensity profiles are plotted in Figs. \ref{FigS6}B and \ref{FigS6}C (in red).  We used two fitting protocols for these results.  For SD, below $z$ = 100 nm, $s_{x,y} = s_0 \sqrt{1+ (z/990.3)^2}$ where $s_0 = 108.2$ nm is the minimum in the eGFP $s_{x,y}$ vs $z$ curve that defines the focus; above $z$ = 100 nm, a linear fit yields a slope of 0.73.  The shape of the experimental defocusing eGFP $s_{x,y}$ vs $z$ curve is consistent with that of the theoretical results, but the values are 30 nm higher.  These higher experimental SD values are consistent with the reported values in recent publications using similar imaging setups to ours \cite{Wang2006,Wang2010,Selvin2003,Selvin2004}, and it is due to a combination of the pixelation effect of the camera, finite bandwidth of the emission filter, inhomogeneity of the molecule's emission polarization, and the imperfection of the current single-molecule imaging systems.  Because of this increase in the  experimental eGFP SD values, the corresponding experimental eGFP PSF amplitudes are lower than the theoretical values.  Below $z$ = 150 nm, $C(z) = \frac{1}{1+(\frac{z+140}{726.7})^2}$; above $z$ = 150 nm, $C(z) = \frac{1}{1+(\frac{z+140}{389.4})^2}$.  The peak of $C(z)$ does not coincide with the SD minimum at $z$ = 0; rather it is shifted to -140 nm.  The experimental eGFP functions were used for theoretical and simulation diffusing eGFP  SD studies in this article.\vspace{1mm}\\
\textbf{Mean emitted photon counts at each simulation step.}\\
The number of photons emitted at each simulation step is a random number drawn from a Poisson distribution with a mean value being $Aexp(-z/d)C(z)s(z)^2$, where $exp(-z/d)$ describes the exponentially decaying evanescent light intensity and $d \approx$ 117 nm is the penetration depth calculated according to our incident angle of 70$^{\circ}$ \cite{Axelrod1989}, $\frac{C(z)}{2{\pi}}$ is the amplitude of the refractive-index-mismatch affected eGFP PSF, $2{\pi}s(z)^2$ corrects for the amplitude ($\frac{1}{2{\pi}s(z)^2}$ )of a simulated Gaussian PSF with SD $s(z)$, and $A$ is a scaling factor that accounts for the quantum efficiency of eGFP molecules.  Below we obtain $C(z)$ and $A$.

We first describe how $C(z)$ is used.  When we simulate a PSF by distributing $N$ photons following a 2D Gaussian spatial distribution with SD $s(z)$, the amplitude of the PSF will be $\frac{N}{2{\pi}s(z)^2}$, where $\frac{1}{2{\pi}s(z)^2}$ is the amplitude of a normalized Gaussian function with SD $s(z)$ for one photon.  However, the Gaussian PSFs with $N$ photons in the refractive-index-mismatch case have the same SD $s(z)$ but a different amplitude, $\frac{C(z)N}{2{\pi}}$.  Thus, when simulating the refractive-index-mismatch affected PSFs by spatially distributing photons using a 2D Gaussian distribution with SD $s(z)$, each photon count should be corrected by factor $C(z)s(z)^2$,  where $s(z)^2$ cancels the amplitude of the simulated normalized PSF.   

To determine $A$, a random value was picked to simulate PSF photon distributions at a finite exposure time.  After the photon to camera count conversion (with a conversion factor $M$ = 1 which introduces additional variance to the emission photon distributions \cite{Wang2010}), the modified emission photon count distributions were compared to the experimental distributions at the same exposure times. $A$ was obtained when a good match between the two distributions was achieved.  Figure S\hspace{-1mm} \ref{FigS3} compares the 0.6 ms experimental and simulated photon emission distributions; $A$ was 0.80 and remains approximately constant for all exposure times (the mean $A$ value for the 0.3 ms to 1 ms exposure times is 0.86; data not shown).
 
In the theoretical $s_x$ calculations using Eq. S\ref{Defocus}, $A$ was not included since it does not affect the final calculated PSF $s_x$ results.\vspace{1mm}\\
\textbf{Starting locations of the imaged diffusing eGFP molecules.}\\
In order to correctly simulate diffusing eGFP molecules near fused-silica surfaces, the axial starting positions are needed.  We obtained the eGFP diffusion starting position probability distributions at different exposure times by simulating a fluorophore's emitted photon distributions for a range of starting positions.

At each exposure time, we simulated 1000 axial-direction diffusion trajectories starting from the fused-silica-water interface to an extended distance in water for the exposure time ($z$ = 0 nm to $117 + 3\sqrt{2D_3t}$ nm measuring from the fused-silica surface at focus).  A reflective fused-silica surface at $z$ = 0 was used.  The simulations included the triplet state effect, and the number of photons emitted at each step is described above with the difference of using the mean of the Poisson  emission photon count distribution at each step, rather than drawing a random number from the Poisson distribution.  At each starting position, we obtained the ratio of the number of photons emitted within the penetration depth ($d = 117$ nm) to all emitted photons for a simulated trajectory, and then the mean ratio for all simulated 1000 trajectories was obtained and plotted in Fig. S\ref{FigS2}.  The exposure times shown are 0.3, 0.7, and 1 ms, and half-Gaussian functions are fit to the distributions.  The fitted SD values of the starting position distribution functions increase with the exposure time as $SD(t) = 1.538\times{10^5}t+122.4$ nm.  For $t$ = 0.3 ms, Fig. S\ref{FigS2}A shows that most molecules we observed experimentally should start within 200 nm of the surface.\vspace{1mm}\\  
\textbf{Axial-direction PWDF$_z$.}\\
In order to obtain the eGFP PWDF in the axial direction, we performed axial-direction diffusion simulations for all exposure times using the starting position distributions described above and a reflective fused-silica surface at $z$ = 0.  Figure S\hspace{-1mm} \ref{FigS7}A shows 9 representative simulated PWDF$_z$s for the 0.6 ms exposure time.  Since 84.5$\%$ of the data can be fit by a Gaussian function with $R^2 > 0.7$, we use Gaussian function to approximate PWDF$_z$s in Eq. S\ref{Defocus}.  Figure S\hspace{-1mm} \ref{FigS7}B shows the SD distribution of the fitted PWDF$_z$s and the Gaussian fit to the distribution; the mean SD = $\sqrt{A_z\cdot2D_3t}$ is 75.8 nm, yielding $A_z$ = 0.054.  $A_z$ remains constant for other exposure times with a mean value of 0.052.   Figure S\hspace{-1.8mm} \ref{FigS7}C shows the mean value ($z_0$) distribution of the fitted PWDF$_z$s and the Gaussian fit to the distribution; the mean $z_0$ is 142.7 nm.  The inset in Fig. S\ref{FigS7}C shows that $\langle{z_0}\rangle$ increases with $t$ as $\langle{z_0}\rangle = \sqrt{0.27D_3t}$ + 25.5 nm.  For each exposure time, 1000 trajectories were simulated to obtain the results.\\\\

\newpage
\bibliographystyle{Science}

\begin{thebibliography}{10}

\bibitem{Wang2006}
Y.~M. Wang, R.~H. Austin, and Edward~C. Cox.
\newblock Single molecule measurements of repressor protein 1\textsc{D}
  diffusion on \textsc{DNA}.
\newblock {\em Physical Review Letters}, 97:048302, 2006.

\bibitem{Greene2008}
Jason Gorman and Eric~C Greene.
\newblock Visualizing one-dimensional diffusion of proteins along dna.
\newblock {\em Nature Structural and Molecular Biology}, 15:768--774, 2008.

\bibitem{Moerner2010}
Michael~A. Thompson, Matthew~D. Lew, Majid Badieirostami, and W.~E. Moerner.
\newblock Localizing and tracking single nanoscale emitters in three dimensions
  with high spatiotemporal resolution using a double-helix point spread
  function.
\newblock {\em Nano Letters}, 10:211--218, 2010.

\bibitem{Stoeckli1998}
Manfred Auer, Keith~J. Moore, Franz~J. Meyer-Almes, Rolf Guenther, Andrew~J.
  Pope, and Kurt~A. Stoeckli.
\newblock Fluorescence correlation spectroscopy: lead discovery by miniaturized
  \textsc{HTS}.
\newblock {\em Drug Discoveries and Therapeutics}, 3(10):457--465, 1998.

\bibitem{Brock2009}
Alexander Ganser, Gnter Roth, Joost~C. van Galen, Janneke Hilderink, Joost
  J.~G. Wammes, Ingo Mller, Frank~N. van Leeuwen, Karl-Heinz Wiesmller, and
  Roland Brock*†.
\newblock Diffusion-driven device for a high-resolution dose−response
  profiling of combination chemotherapy.
\newblock {\em Analytical Chemistry}, 81:5233--5240, 2009.

\bibitem{Niemeyer2003}
Lizhong He and Bernd Niemeyer.
\newblock A novel correlation for protein diffusion coefficients based on
  molecular weight and radius of gyration.
\newblock {\em Biotechnology Progress}, 19:544--548, 2003.

\bibitem{NMR2005}
G.~A. Webb.
\newblock {\em Nuclear magnetic resonance}, volume~34.
\newblock The Royal Society of Chemistry, Thomas Graham House, Science Park,
  Milton Road, Cambridge CB40WF, UK, 2005.

\bibitem{DLS2000}
B.J. Berne and R.~Pecora.
\newblock {\em Dynamic Light Scattering: With application to Chemistry, Biology
  and Physics.}
\newblock General Publishing Company, Ltd, 30 Lesmill Road, Don Mills, Toronto,
  Ontario, 2000.

\bibitem{Webb1972}
Douglas Magde, Elliot Elson, and W.~W. Webb.
\newblock Thermodynamic fluctuations in a reacting system - measurement by
  fluorescence correlation spectroscopy.
\newblock {\em Physical Review Letters}, 29:705--708, 1972.

\bibitem{Nienhaus2004}
Andreas Schenk, Sergey Ivanchenko, Carlheinz R$\ddot{o}$cker, J$\ddot{o}$rg
  Wiedenmann, and G.~Ulrich Nienhaus.
\newblock Photodynamics of red fluorescent proteins studied by fluorescence
  correlation spectroscopy.
\newblock {\em Biophysical Journal}, 86:384--394, 2004.

\bibitem{Schwille2008}
Zden$\check{e}$k Petr$\acute{a}$$\check{s}$ek and Petra Schwille.
\newblock Precise measurement of diffusion coefficients using scanning
  fluorescence correlation spectroscopy.
\newblock {\em Biophysical Journal}, 94:1437--1448, 2008.

\bibitem{Wiersma2001}
Eric~O. Potma, Wim~P. de~Boeij, Leonard Bosgraaf, Jeroen Roelofs, Peter J.~M.
  van Haastert, and Douwe~A Wiersma.
\newblock Reduced protein diffusion rate by cytoskeleton in vegetative and
  polarized dictyostelium cells.
\newblock {\em Biophysical Journal}, 81:2010--2019, 2001.

\bibitem{Xie2006}
P.~C. Blainey, A.~M. van Oijen, A.~Banerjee, G.~L. Verdine, and X.~S. Xie.
\newblock A base-excision \textsc{DNA}-repair protein finds intrahelical lesion
  bases by fast sliding in contact with \textsc{DNA}.
\newblock {\em Proceedings of the National Academy of Sciences of the United
  States of America}, 103(15):5752--5757, 2006.

\bibitem{Nachury2010_2}
M.~V. Nanchury, E.~S. Seeley, and Hua Jin.
\newblock Trafficking to the cialiary membrane: How to get across the
  periciliary diffusion barrier?
\newblock {\em Annual Review of Cell and Developmental Biology}, 26:59--87,
  2010.

\bibitem{Spector2007}
Ken Ritchie and Jeff Spector.
\newblock Single molecule studies of molecular diffusion in cellular membranes:
  determining membrane structure.
\newblock {\em Biopolymers}, 87:95--101, 2007.

\bibitem{Qian1991}
Hong Qian, Michael~P. Sheetz, and Elliot~L. Elson.
\newblock Single particle tracking: Analysis of diffusion and flow in
  two-dimensional systems.
\newblock {\em Biophysical Journal}, 60:910--921, 1991.

\bibitem{Gratton2005}
V.~Levi, Q.~Ruan, and E.Gratton.
\newblock 3-\textsc{D} particle tracking in a two-photon microscope:
  Application to the study of molecular dynamics in cells.
\newblock {\em Biophysical Journal}, 88:2919--2928, 2005.

\bibitem{Schmidt2011}
Stefan Semrau, Anna Pezzarossa, and Thomas Schmidt.
\newblock Microsecond single-molecule tracking ($\mu$s\textsc{SMT}).
\newblock {\em Biophysical Journal}, 100:L19--L21, 2011.

\bibitem{Wang2010_2}
Shawn~H. DeCenzo, Michael~C. DeSantis, and Y.~M. Wang.
\newblock Single-image separation measurements of two unresolved fluorophores.
\newblock {\em Optics Express}, 18(16):16628--16639, 2010.

\bibitem{Borczyskowski2002}
J.~Schuster, F.~Cichos, and C.~von Borczyskowski.
\newblock Diffusion measurements by single-molecule spot-size analysis.
\newblock {\em The Journal of Physical Chemistry A}, 106(22):5403--5406, June
  2002.

\bibitem{Wang2011_2}
Shanen Kian~G. Zareh and Y.~M. Wang.
\newblock Single-molecule imaging of protein-surface adsorption mechanisms.
\newblock {\em Microscropy Research and Technique}, 74:682--687, 2011.

\bibitem{Wang2010}
Michael~C. DeSantis, Shawn~H. DeCenzo, Je-Luen Li, and Y.~M. Wang.
\newblock Precision analysis for standard deviation measurements of immobile
  single fluorescent molecule images.
\newblock {\em Optics Express}, 18(7):6563--6576, March 2010.

\bibitem{Verkman1997}
R.~Swaminathan, Cathy~P. Hoang, and A.~S. Verkman.
\newblock Photobleaching recovery and anisotropy decay of green fluorescent
  protein \textsc{GFP-S}65\textsc{T} in solution and cells: Cytoplasmic
  viscosity probed by green fluorescent protein translational and rotational
  diffusion.
\newblock {\em Biophysical Journal}, 72:1900--1907, 1997.

\end{thebibliography}

\begin{thebibliography}{10}

\bibitem{Wang2011_2}
Shanen Kian~G. Zareh and Y.~M. Wang.
\newblock Single-molecule imaging of protein-surface adsorption mechanisms.
\newblock {\em Microscropy Research and Technique}, 74:682--687, 2011.

\bibitem{Wang2010}
Michael~C. DeSantis, Shawn~H. DeCenzo, Je-Luen Li, and Y.~M. Wang.
\newblock Precision analysis for standard deviation measurements of immobile
  single fluorescent molecule images.
\newblock {\em Optics Express}, 18(7):6563--6576, March 2010.

\bibitem{Wang2010_2}
Shawn~H. DeCenzo, Michael~C. DeSantis, and Y.~M. Wang.
\newblock Single-image separation measurements of two unresolved fluorophores.
\newblock {\em Optics Express}, 18(16):16628--16639, 2010.

\bibitem{Briddon2004}
K.~Weisshart, V.~J$\ddot{u}$ngel, and S.~J. Briddon.
\newblock The \textsc{LSM} 510 \textsc{META} - \textsc{C}onfo\textsc{C}or 2
  system: An integrated imaging and spectroscopic platform for single-molecule
  detection.
\newblock {\em Current Pharmaceutical Biotechnology}, 5:135--54, 2004.

\bibitem{Schwille2008}
Zden$\check{e}$k Petr$\acute{a}$$\check{s}$ek and Petra Schwille.
\newblock Precise measurement of diffusion coefficients using scanning
  fluorescence correlation spectroscopy.
\newblock {\em Biophysical Journal}, 94:1437--1448, 2008.

\bibitem{Hell2006}
Katrin~I Willig, Robert~R. Kellner, Rebecca Medda, Birka Hein, STefan Jakobs,
  and Stefan~W Hell.
\newblock Nanoscale resolution in \textsc{GFP}-based microscopy.
\newblock {\em Nature Methods}, 3(9):721--723, 2006.

\bibitem{Hell1999}
A.~Egner and S.~W. Hell.
\newblock Equivalence of the \textsc{H}uygens--\textsc{F}resnel and debye
  approach for the calculation of high aperture point-spread functions in the
  presence of refractive index mismatch.
\newblock {\em Journal of Microscopy}, 193:244--249, 1999.

\bibitem{ZhuangCylinder2008}
Bo~Huang, Wenqin Wang, Mark Bates, and Xiaowei Zhuang.
\newblock Three-dimensional super-resolution imaging by stochastic optical
  reconstruction microscopy.
\newblock {\em Science}, 319:810--813, 2008.

\bibitem{Moerner2010}
Michael~A. Thompson, Matthew~D. Lew, Majid Badieirostami, and W.~E. Moerner.
\newblock Localizing and tracking single nanoscale emitters in three dimensions
  with high spatiotemporal resolution using a double-helix point spread
  function.
\newblock {\em Nano Letters}, 10:211--218, 2010.

\bibitem{Wang2006}
Y.~M. Wang, R.~H. Austin, and Edward~C. Cox.
\newblock Single molecule measurements of repressor protein 1\textsc{D}
  diffusion on \textsc{DNA}.
\newblock {\em Physical Review Letters}, 97:048302, 2006.

\bibitem{Selvin2003}
Ahmet Yildiz, Joseph~N. Forkey, Sean~A. McKinney, Taekjip Ha, Yale~E. Goldman,
  and Paul~R. Selvin.
\newblock Myosin \textsc{V} walks hand-over-hand: \textsc{S}ingle fluorophore
  imaging with 1.5-nm localization.
\newblock {\em Science}, 300:2061--2065, 2003.

\bibitem{Selvin2004}
Ahmet Yildiz, Michio Tomishige, Ronald~D. Vale, and Paul~R. Selvin.
\newblock Kinesin walks hand-over-hand.
\newblock {\em Science}, 303:676--678, 2004.

\bibitem{Axelrod1989}
D.~Axelrod.
\newblock Total internal-reflection fluorescence microscopy.
\newblock {\em Methods in Cell Biology}, 30:245--270, 1989.

\end{thebibliography}

\newpage

\begin{figure}
\includegraphics[width=5in]{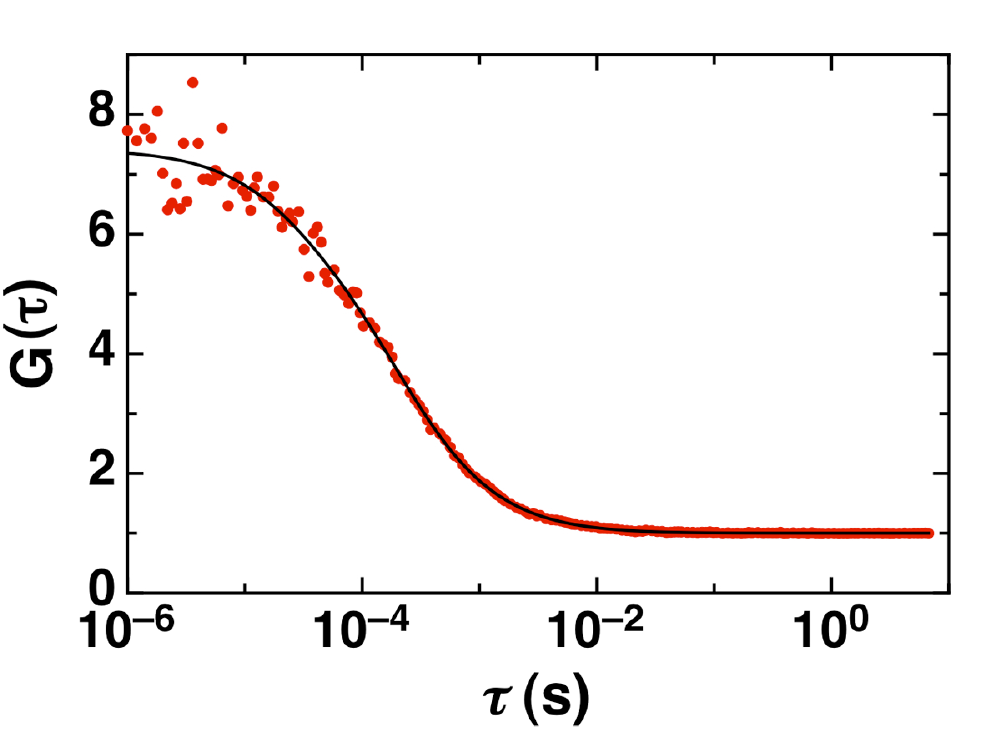}
\caption{Diffusing eGFP FCS autocorrelation plot.  The black curve is a fit to the raw data (red dots).}
\label{FigS1}
\end{figure}

\begin{figure}
\includegraphics[width=4in]{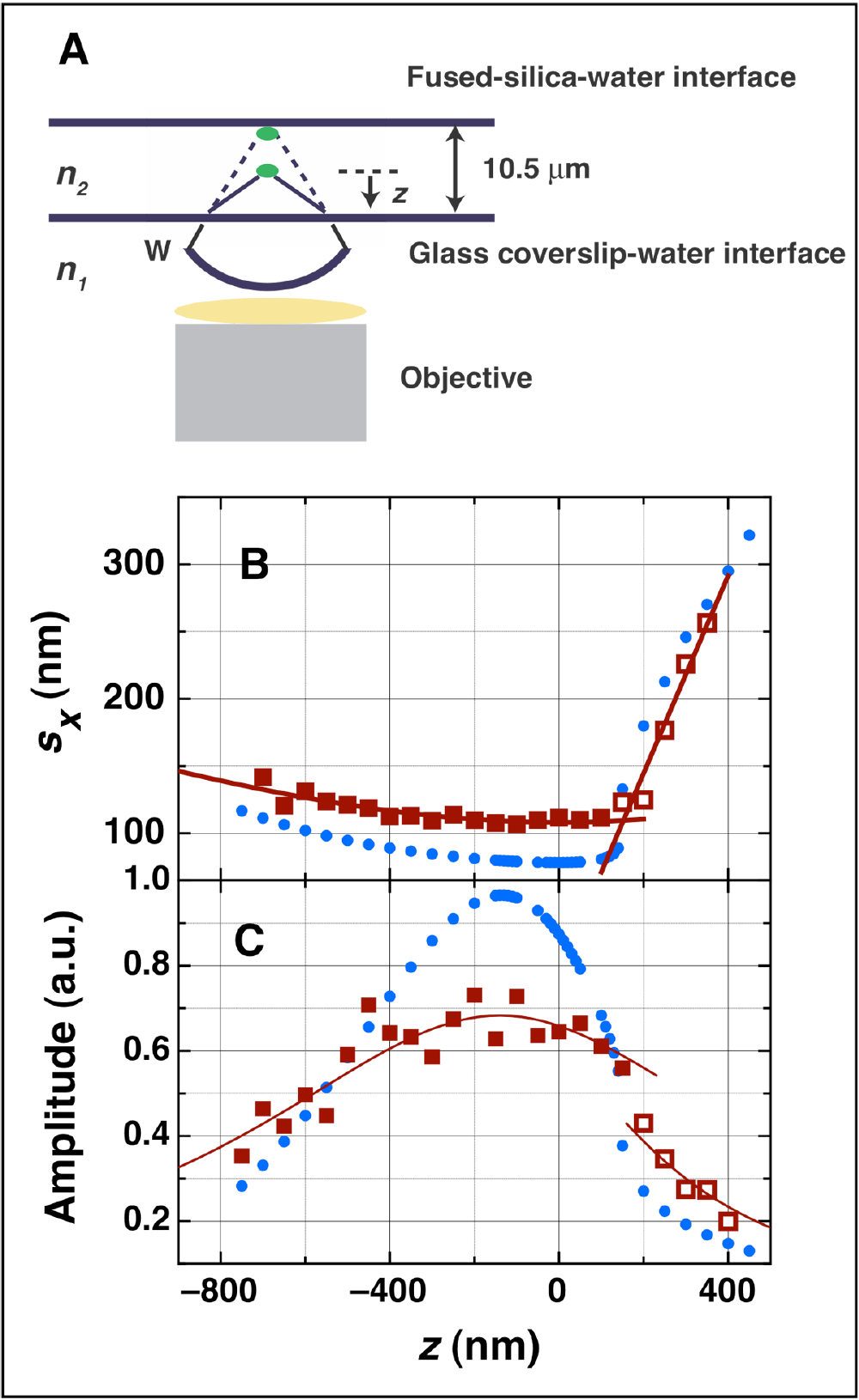}
\caption{(A) Our imaging setup and the schematics of emission from a fluorophore in water located at the fused-silica-water interface 10.5 $\mu$m away from the coverslip-water interface.  Dashed lines trace the emission from the actual location of the fluorophore, and solid lines trace the emission from the apparent location of the fluorophore.  The letter ``W" labels the wavefront of the emission before reaching the objective.  (B) Calculated (blue) and experimental (red) eGFP PSF SDs and (C) amplitude vs the defocusing distance $z$ plots.  Lines are fits to the experimental measurements.  The focal point is at the minimum of the eGFP $s_{x}$ vs $z$ curve (same for $s_y$).}
\label{FigS6}
\end{figure}

\begin{figure}
\includegraphics[width=5in]{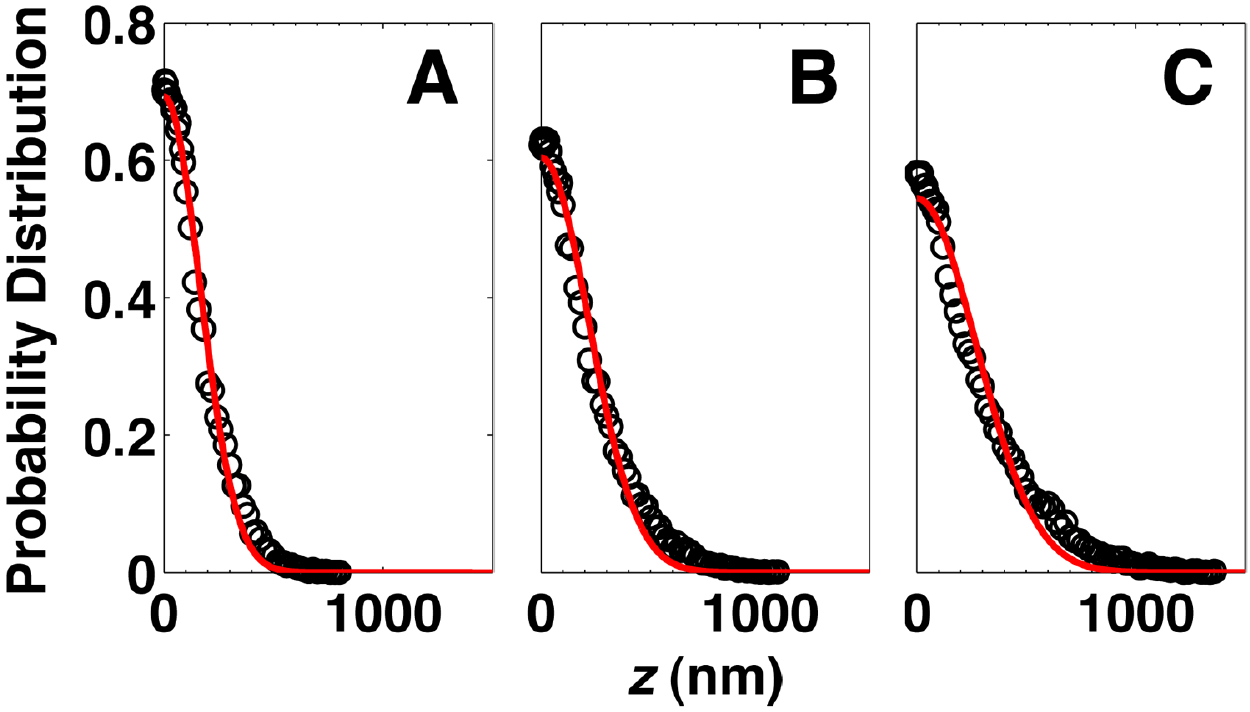}
\caption{Simulation results for the diffusing eGFPs' starting-location distribution near the fused-silica-water interface at exposure times of 0.3, 0.7, and 1ms (A, B, and C) and their corresponding fits.}
\label{FigS2}
\end{figure}

\begin{figure}
\includegraphics[width=5in]{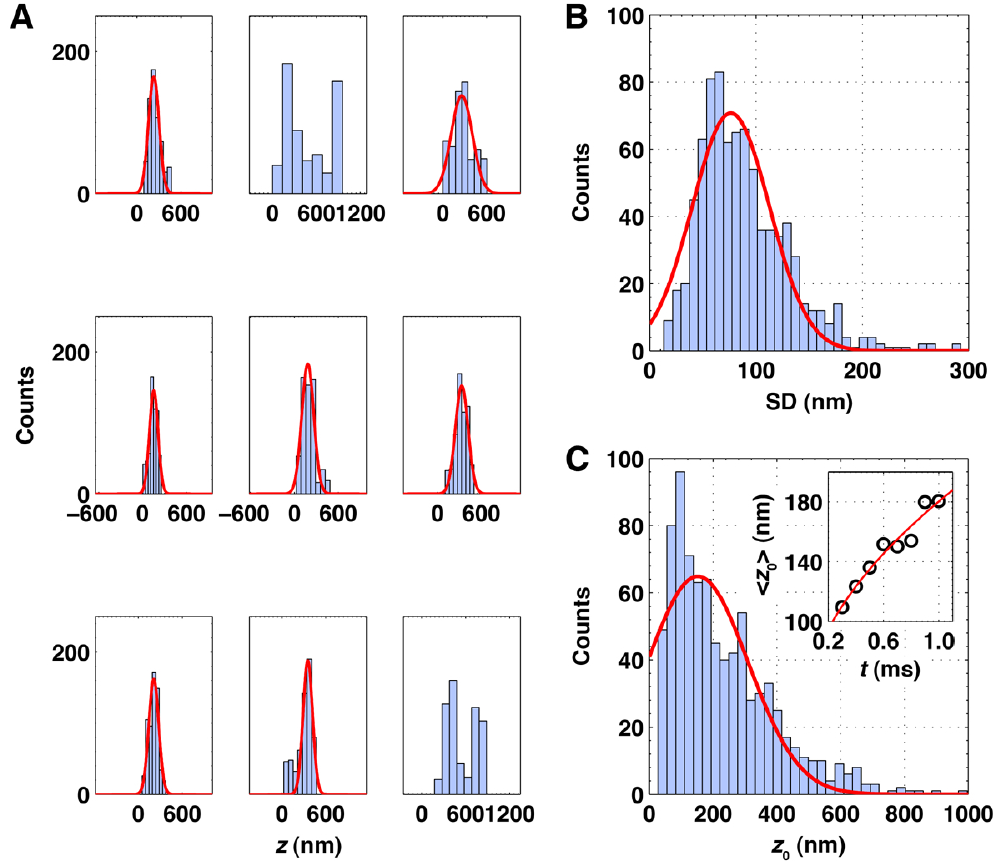}
\caption{Study of the axial-direction PWDF parameters.  (A) Nine randomly selected PWDF$_z$s at $t$ = 0.6 ms.  (B) Fitted-PWDF$_z$s' SD distribution and its Gaussian fit.  The mean is 75.8 nm.  (C) Fitted-PWDF$_z$s' mean ($z_0$) distribution and the Gaussian fit.  Inset, $\langle{z_0}\rangle$ increases with $D_3t$.}
\label{FigS7}
\end{figure}

\begin{figure}
\includegraphics[width=5in]{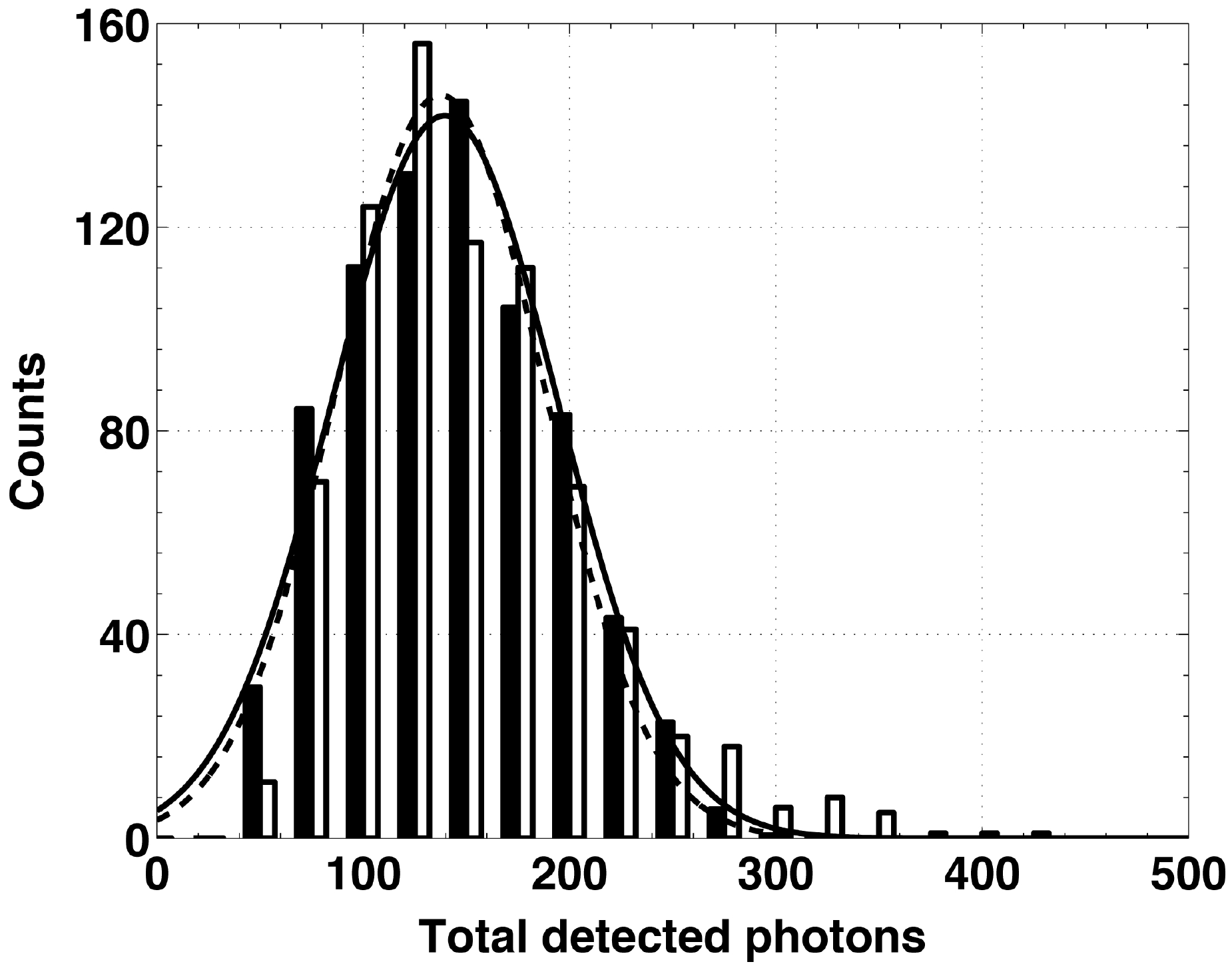}
\caption{Comparing the experimental (black) and simulated (empty) photon emission distributions at 0.6 ms.  Their respective Gaussian fits in solid and dashed lines are in good agreement.}
\label{FigS3}
\end{figure}

\begin{figure}
\includegraphics[width=5in]{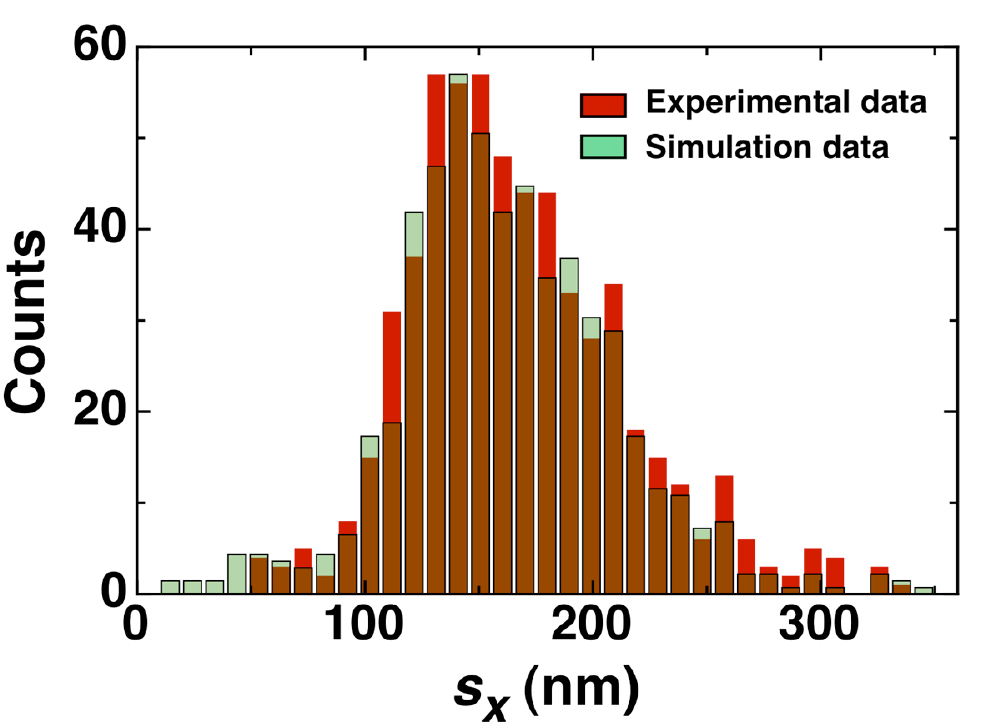}
\caption{Comparing the experimental (green) and simulated (red) diffusing eGFP SD distributions.}
\label{FigS5}
\end{figure}


\end{document}